\documentclass[twocolumn,aps,prd,longbibliography, replace]{revtex4-1}
\usepackage{graphicx,bm,xcolor, bbold}
\usepackage[normalem]{ulem}
\usepackage{hyperref}
\hypersetup{colorlinks, linkcolor={blue},citecolor={blue},urlcolor={blue}}
\usepackage{physics}
\usepackage[utf8]{inputenc}
\usepackage[english]{babel}
\usepackage{tikz}
\usepackage{amsmath}
\makeatletter
\def\graphicscale{\twocolumn@sw{0.3}{0.4}}
\def\graphicthreescale{\twocolumn@sw{0.3}{0.4}}

\usepackage{dsfont}
\usepackage{scalerel}
\usepackage{graphicx}
\usepackage{dcolumn}
\usepackage{bm}
\usepackage{amsmath}
\usepackage{braket}
\usepackage{epstopdf}
\usepackage{placeins}
\usepackage{multirow}
\usepackage{makecell}
\usepackage{cleveref}
\usepackage{xcolor}
\usepackage{dblfloatfix}
\usepackage{ulem}
\usepackage{comment}

\newcommand{\beq}{\begin{eqnarray}}
\newcommand{\eeq}{\end{eqnarray}}
\newcommand{\beqnn}{\begin{eqnarray*}}
\newcommand{\eeqnn}{\end{eqnarray*}}

\newcommand{\lin}{\mathrm{lin}}

\usepackage[font=small, labelfont=bf, textfont={small,it}]{caption}

\def\spose#1{\hbox to 0pt{#1\hss}}
\def\ltapprox{\mathrel{\spose{\lower 3pt\hbox{$\mathchar"218$}}
	\raise 2.0pt\hbox{$\mathchar"13C$}}}

\graphicspath{{figs/}}

\begin{document}

\title{Neural Networks Asymptotic Behaviours for the Resolution of Inverse Problems}

\author{Luigi Del Debbio$^a$}
\author{Manuel Naviglio$^{b,c}$}
\author{Francesco Tarantelli$^b$}
\email{manuel.naviglio@sns.it}
\email{francesco.tarantelli@phd.unipi.it}
\affiliation{$^a$Higgs Centre for Theoretical Physics, School of Physics and Astronomy, The University of Edinburgh, Edinburgh EH9 3FD, UK}
\affiliation{$^b$Dipartimento di Fisica dell'Universit\`a di Pisa \& INFN Sezione di Pisa, Largo Pontecorvo 3, I-56127 Pisa, Italy}
\affiliation{$^c$Scuola Normale Superiore, Piazza dei Cavalieri 7, 56126 Pisa, Italy}

\date{\today}
\begin{abstract}
This paper presents a study of the effectiveness of Neural Network (NN) techniques for deconvolution inverse problems relevant for applications in Quantum Field Theory, but also in more general contexts. We consider NN’s asymptotic limits, corresponding to Gaussian Processes (GPs), where non-linearities in the parameters of the NN can be neglected. Using these resulting GPs, we address the deconvolution inverse problem in the case of a quantum harmonic oscillator simulated through Monte Carlo techniques on a lattice. In this simple toy model, the results of the inversion can be compared with the known analytical solution. Our findings indicate that solving the inverse problem with a NN yields less performing results than those obtained using the GPs derived from NN’s asymptotic limits. Furthermore, we observe the trained NN’s accuracy approaching that of GPs with increasing layer width. Notably, one of these GPs defies interpretation as a probabilistic model, offering a novel perspective compared to established methods in the literature. 
Our results suggest the need for detailed studies of the
training dynamics in more realistic set-ups.
\end{abstract}

\maketitle
\section{Introduction}
Inverse problems are a wide class of problems affected by relevant mathematical and 
numerical issues. These problems, where a continuous functions needs to be reconstructed 
from a finite set of data, are central in many fields of research. Intuitively, solving 
inverse problems corresponds to finding a transformation that yields a gain of information, 
making the problem ill-posed for a continuous function, and ill-conditioned when the
function is discretised. In this study, we focus on a correlation function $C(\tau)$, 
whose spectral decomposition yields
\begin{equation}
    \label{eq: SpectralRelation}
    C(\tau) = \int_0^{\infty} d\omega \rho(\omega) b(\omega, \tau)\, ,
\end{equation}
where $\rho(\omega)$ is the spectral density of the 
theory, and $b(\omega,\tau)$ a known function. Solving the inverse problem
amounts to extracting information about the spectral function from a discrete set of 
measurements of $C(\tau)$.
The ill-posedness of these problems implies that it is not possible to find a unique, 
exact solution. Instead the final answer will depend on the assumption that are 
made in solving the problem. 
The Backus-Gilbert (BG) method~\cite{Backus_Gilbert,Hansen:2017mnd,Hansen:2019idp}, 
reviewed in Appendix~\ref{BG review}, has been widely used in lattice field theory in 
recent
years~\cite{Brandt:2015sxa,Hansen:2017mnd,Bulava:2019kbi,Frezzotti:2023nun,Almirante:2023wmt,Bonanno:2023ljc,Bonanno:2023thi,Bonanno:2023xfv}.\\
Another possible approach to solve inverse problems is based on Gaussian Processes 
(GP)~\cite{Horak:2021syv,DelDebbio:2021whr}. It has been demonstrated that BG and GP 
methods are analytically equivalent~\cite{10.1093/gji/ggz520}, see e.g. 
Ref.~\cite{DelDebbio:2023pyg} for a recent application to determinations of the 
spectral density. We refer the reader to Appendix~\ref{BG review} for a summary of the 
relevant details. 

Recently, there have been multiple studies dedicated to the extraction of
spectral densities using Neural Networks (NN) 
techniques~\cite{Kades:2019wtd,laanait2019exascale,Chen:2021giw,Wang:2021jou,Wang:2021cqw,Shi:2022yqw,Lechien:2022ieg,Boyda:2022nmh,Buzzicotti:2023qdv}. 
However, the efficacy of Neural Network techniques in yielding substantial improvements 
over traditional methods needs to be critically scrutinised, as it is unclear 
whether these 
techniques yield appreciable corrections compared to conventional approaches such 
as BG and GP, or if their utilization is somehow redundant.
A notable study has been presented in Ref.~\cite{Buzzicotti:2023qdv}, where the authors 
compared the results obtained from a model independent training using a Convolutional 
Neural Network with the ones obtained from the BG approach. The overall conclusion of the 
study seems to suggest an agreement among these two results.

In this paper, we apply theoretical results from computer science to explore 
numerically the asymptotic limits of neural networks applied to the solution of 
the inverse problem~\eqref{eq: SpectralRelation}, in a case where the spectral 
densities are know, namely the quantum-mechanical harmonic oscillator. 
Our investigation focuses on the connection 
between these limits and the GP techniques, in the specific case where the output of 
the NN is a linear function of its parameters. These limits allow an analytic expression 
for the output of the NN at the end of the training process, thereby simplifying the 
study of the training and the subsequent test results. Moreover, we analyse whether 
the non-linear corrections, which appear when considering neural networks of finite width, 
are relevant in the resolution of these inverse problems. 

The paper is organised as follows. In Section~\ref{Asymptotic} we discuss two asymptotic 
limits of neural networks, infinite width and linear limit, emphasising their interplay. 
Our conventions and some basic results about Neural Networks can be found in 
Appendix~\ref{NN_Intro}. In Section~\ref{Equivalence}, we highlight the equivalence of the 
asymptotic limits of NNs with GP methods. Finally, in Section~\ref{Experiments}, we perform 
numerical experiments using lattice simulations of the quantum-mechanical harmonic 
oscillator, whose spectral function is analytically known, in which we compare the results 
obtained using a finite NN with the ones obtained using the GPs resulting as the asymptotic 
limit of the NN.

\section{Neural Network asymptotic limits}\label{Asymptotic}
In Appendix~\ref{NN_Intro}, we describe the fundamental properties of fully connected neural networks under gradient descent (GD) training.
In this section, we show that equations~\eqref{eq: GradDesc} simplify in suitable limits of the architecture of the NN. We describe two fundamental limits of neural networks: the linear limit and the infinite width limit, initially independently, and subsequently highlighting their correspondences under various conditions.

Let us introduce here the notation that we use in the rest of the paper. 
The input vector of the NN is made of the values of a correlator $\mathcal{C}$ 
at $n_0$ Euclidean 
times. We will use the symbol $\tau$ to denote the times at which the correlators are 
computed; these should not be confused with training time $t$. In the notation of 
Appendix~\ref{NN_Intro}, we have for the input vector,
\begin{equation}
    \vec{x} = 
    \begin{pmatrix}
        \mathcal{C}(\tau_1)\\
        \mathcal{C}(\tau_2)\\
        \ldots \\
        \mathcal{C}(\tau_{n_0})
    \end{pmatrix}\, ,
\end{equation}
with $n_0=100$. The output of the NN yields the value of the spectral density
at $n_L=1000$ values of the energy $\omega$, 
\begin{equation}
     \rho(\omega_i) = \phi^{(L)}_i(\mathcal{C})\, , \quad i=1,\ldots,n_L\, .
\end{equation}

The training set $\mathcal{A} = \left\{(\mathcal{C}_\alpha, \varrho_\alpha)\right\}$ is made of pairs of correlators and their corresponding spectral density, each pair 
being labelled by an index $\alpha$. Therefore, each value of $\alpha$ identifies one 
input of the NN, 
\begin{equation}
    \mathcal{C}_\alpha = 
    \begin{pmatrix}
        \mathcal{C}_\alpha(\tau_1)\\
        \mathcal{C}_\alpha(\tau_2)\\
        \ldots \\
        \mathcal{C}_\alpha(\tau_{n_0})
    \end{pmatrix}\, ,
\end{equation}
and the spectral density for that correlator, represented by the vector
\begin{equation}
    \varrho_\alpha = 
    \begin{pmatrix}
        \rho_\alpha(\omega_1)\\
        \rho_\alpha(\omega_2)\\
        \ldots \\
        \rho_\alpha(\omega_{n_L})\\
    \end{pmatrix}\, .
\end{equation}
In what follows we denote $\boldsymbol{\mathcal{C}}$ and $\boldsymbol{\mathcal{\varrho}}$ the sets of input correlators and densities, and $\bar{\rho}$ the output of the NN for a generic input correlator $\bar{\mathcal{C}}$, 
\begin{equation}
    \bar{\rho}_i = \phi^{(L)}_i(\bar{\mathcal{C}})\, .
\end{equation}
We refer to $\bar{\rho}$ as the output of the NN and $\bar{C}$ as a generic test point. 

\subsection{The linear limit}
In the context of neural networks, the linear limit refers to a scenario where the NN's behaviour becomes predominantly linear in the parameters. This simplification allows for a more interpretable and analytically tractable understanding of the NN's behaviour, making it a valuable tool for theoretical analysis and certain applications in machine learning~\cite{lee2019wide}.\\
This limit holds when the output $\bar{\rho}(\bar{C},\theta_t)$ of the NN can be approximated by the following first order Taylor expansion 
\begin{equation}
\bar{\rho}^\lin(\bar{C}, \theta_t) \equiv \bar{\rho}(\bar{C},\theta_t)|_{\theta_t=\theta_0} + \nabla_\theta \bar{\rho}(\bar{C}, \theta_t)|_{\theta_t = \theta_0}\omega_t
\end{equation}
where the vector $\theta_t$ indicates the vector of all the parameters of the network at the time $t$, $\omega_t = \theta_t-\theta_0$ the variation of the parameters from their initialisation value, while $\bar{\rho}(\bar{C},\theta_0)$ is the output of the neural network after the initialisation of the parameters.

When applying the linear approximation to the NN parameters, the NN's output remains a nonlinear function of its input due to activation functions and architecture. However, this approximation sacrifices intrinsic nonlinearities in the parameters, limiting adaptation to complex data. This constraint hinders the NN's ability to represent complex nonlinear behaviours arising from interactions between parameters.

If this approximation holds, Eqs.~\eqref{eq: GradDesc} become analytically solvable; the dynamics of the parameters and of the output of the NN are given by
\begin{equation}
\begin{split}
    & \omega_t = - \sum_{\alpha_1,\alpha_2} \nabla_\theta \rho(\mathcal{C}_{\alpha_1},\theta_0) \left[\hat{\Theta}_0 \bigg(\mathcal{I} - e^{-\eta \hat{\Theta}_0t}\bigg)\right]_{\alpha_1\alpha_2} \\ 
    & \phantom{\omega_t = -} \times \big(\rho(\mathcal{C}_{\alpha_2},\theta_0) - \varrho_{\alpha_2}\big),\\
    & \rho^\lin(\mathcal{C}_\alpha, \theta_t) = \big( \mathcal{I} - e^{-\eta \hat{\Theta}_0t}\big)\varrho_\alpha + e^{-\eta\hat{\Theta}_0 t}f(\mathcal{C}_\alpha, \theta_0),
\end{split}
\end{equation}
where $\hat{\Theta}_0$ is the so-called Neural Tangent Kernel (NTK), defined in Eq.~\eqref{eq:NTKdef}, computed at initialization and $\eta$ is the learning rate. Note that the only dependence on the training time is contained in the exponential. As $t$ approaches infinity, the output of the linearized network, denoted as $\bar{\rho}^\lin$, converges to the target output data set $\boldsymbol{\varrho}$ when the training data $\boldsymbol{\mathcal{C}}$ is used as input. Similarly, if training is disregarded, i.e., at $t=0$, the parameters remain unchanged ($\theta_t = \theta_0$), resulting in $\omega_t = 0$.

The output for a generic point $\bar{\mathcal{C}}_\delta$, outside the training set, is
\begin{equation}
\begin{split}
    \label{eq: LinearSol}
    \bar{\rho}^\lin_t(\bar{\mathcal{C}}_\delta,\theta) & =   \bar{\rho}(\bar{\mathcal{C}}_\delta, \theta_0) + \sum_{\alpha} \left[\hat{\Theta}_0\, \hat{\Theta}_0^{-1}\times\right. \\ 
    & \left. \times \bigg(\mathcal{I} - e^{-\eta \hat{\Theta}_0t}\bigg)\right]_{\delta\alpha}\, 
    \big(\varrho_\alpha-\bar{\rho}(\mathcal{C}_\alpha, \theta_0)\big) \\ & 
 = \mu_t(\bar{\mathcal{C}}_\delta) + \gamma_t(\bar{\mathcal{C}}_\delta), 
\end{split}
\end{equation}
where we defined
\begin{equation}
\label{eq: muLin}
    \mu_t(\bar{\mathcal{C}}_\delta) = \left[\hat{\Theta}_0\, \hat{\Theta}_0^{-1}\, \bigg(\mathcal{I} - e^{-\eta \hat{\Theta}_0t}\bigg)
    \right]_{\delta\alpha}
    \varrho_\alpha\, , 
\end{equation}
\begin{align}
    \gamma_t(\bar{\mathcal{C}}_\delta) &= \bar{\rho}(\bar{\mathcal{C}_\delta}, \theta_0) -  \nonumber \\
    & \quad - \left[\hat{\Theta}_0\, \hat{\Theta}_0^{-1} \bigg(\mathcal{I} - e^{-\eta \hat{\Theta}_0t}\bigg)\right]_{\delta\alpha}
    \bar{\rho}(\mathcal{C}_\alpha, \theta_0).
\end{align}
Therefore, if the linear approximation holds  for our specific application, this result provides us with a powerful tool. In fact, we can obtain the output result of the neural network at any given training time t without the need for actual network training. We only need to compute the NTK and the outputs at the initialization and insert them in the formulas.Note that if the training data are already well reconstructed at initialisation , namely $\bar{\rho}(\mathcal{C}_\alpha, \theta_0) = \varrho_\alpha$, then the result is just the output of the NN at initialisation.

\subsection{The large width limit}
\label{LargeWidth}
The infinite width limit of a neural network refers to a theoretical scenario where the number of neurons in each hidden layer is considered to be infinitely large. 

Let us suppose that we initialize the weights matrix and the bias matrix with values taken from a certain probabilistic distribution function, such that the parameters $\theta_0$ are i.i.d and have zero mean and variances $\sigma_w/n^{(\ell)}$ and $\sigma_b$, respectively, with $n^{(\ell)}$ the number of neurons in layer $\ell$. Existing literature \cite{poole2016exponential, schoenholz2017deep, xiao2018dynamical, yang2017mean, lee2018deep, matthews2018gaussian, novak2020bayesian} has established that, as the hidden layer width increases, the output distribution converges to a multivariate Gaussian distribution, with deviations from the Gaussian distribution that scale like powers of $1/n^{(\ell)}$. 
Indeed, each neuron's output is computed as a linear combination of the random i.i.d. weights and biases variables.
When a large number of these random variables enter in the computation of the preactivation functions, according to the Central Limit Theorem, their sum tends to follow a Gaussian distribution, regardless of the original distribution of the individual variables. This means that, as the layers get wider, the overall behaviour of the neural network becomes increasingly Gaussian-like.

This asymptotic behaviour allows the identification of a neural network at initialisation with a Gaussian Process (GP), called NNGP (Neural Network Gaussian Process). In this limit, we can think at the outputs of the NN as sampled from a GP. Furthermore, we are allowed to use a Bayesian inference to discuss the posterior distribution of the neural networks given a set of data, by introducing a kernel function and proceed as routinely done for Gaussian Processes.
Formally, if we consider the output $\bar{\rho}_i(\bar{C},\theta_t)$, where $i=1,...,n^{(L)}$ runs on the dimension of the output of the NN, then in the limit of infinite width we can introduce a kernel function as the covariance matrix of the outputs at initialization in the infinite width limit
\begin{align}
    \Sigma(\bar{C}_{\delta_1},\bar{C}_{\delta_2})_{ij} &= \lim_{n
    \to \infty} \mathbb{E}\bigg[ \bar{\rho}_i(\bar{C}_{\delta_1}) \bar{\rho}_j(\bar{C}_{\delta_2}) \bigg] \\
    &= \delta_{ij} \tilde\Sigma_{\delta_1\delta_2}\, ,
\end{align}
where $n\to\infty$ is a generic notation to denote that all layers become infinitely wide. 
The kernel $\tilde\Sigma_{\delta_1\delta_2}$ can be computed recursively, see~\cite{lee2018deep,lee2019wide} and Appendix~\ref{formulas_sigma_theta}. Thus, given a training input $\boldsymbol{\varrho}$, we have that at initialization, i.e. for $\theta_t=\theta_0$, the output $\bar{\rho}(\boldsymbol{\mathcal{C}},\theta_t)$ will be normally distributed
\begin{equation}
\bar{\rho}(\boldsymbol{\mathcal{C}}, \theta_0) \sim \mathcal{N}(0,\Sigma(\boldsymbol{\mathcal{C}},\boldsymbol{\mathcal{C}}))\, ,
\end{equation}
where we have zero mean since the parameters have zero mean themselves. In order to simplify the notation we have introduced the vector $\bar{\rho}(\boldsymbol{\mathcal{C}},\theta_t)$, such that
\begin{equation}
    \label{eq:VecRhoNotation}
    \bar{\rho}(\boldsymbol{\mathcal{C}},\theta_t)_\alpha = 
    \bar{\rho}(\mathcal{C}_\alpha,\theta_t)\, ,
\end{equation}
and the matrix $\Sigma(\boldsymbol{\mathcal{C}},\boldsymbol{\mathcal{C}})$, such that
\begin{equation}
    \label{eq:MatSigNotation}
    \Sigma(\boldsymbol{\mathcal{C}},\boldsymbol{\mathcal{C}})_{\alpha_1\alpha_2} =
    \Sigma(\mathcal{C}_{\alpha_1},\mathcal{C}_{\alpha_2})\, ,
\end{equation}
and we have omitted the neuron index $i$. 
This result is valid in general for every input $\boldsymbol{\mathcal{C}}$ \textit{at initialization}, since the assumption that the parameters $\theta_t$ are i.i.d. is not necessarily true for $t>0$.\\
This connection suggests that in the case of infinitely wide neural networks, we can substitute an i.i.d. prior on weights and biases with a corresponding GP prior for the output of the nets. As pointed out in Ref.~\cite{NIPS1996_ae5e3ce4}, we can then use Bayesian inference to determine the posterior distribution $\mathcal{P}(\bar{\rho}|\mathcal{D})$, given the data $\mathcal{D}=\{\boldsymbol{\mathcal{C}}, \boldsymbol{\varrho}\}$.
We refer to this procedure as \textit{Bayesian training}: the distribution of weights at initialization determines the Gaussian prior, while Bayes theorem determines the posterior using the training set.  The posterior knowledge about $\bar\rho$ at the test point $\bar{C}$ is encoded in the distribution over the functions $\bar{\rho}$ that are constrained to take values $\boldsymbol{\bar{\varrho}}=(\bar{\varrho}_1, ..., \bar{\varrho}_N)$ over the training inputs $\boldsymbol{\boldsymbol{\mathcal{C}}}=(\boldsymbol{\mathcal{C}}_1, ..., \boldsymbol{\mathcal{C}}_N)$. Here we have introduced $\boldsymbol{\bar{\varrho}}$, which differs from $\boldsymbol{\varrho}$ by some  observational noise as described below. 
Using Bayes theorem,
\begin{equation}
\begin{split}
    \mathcal{P}(\bar{\rho}|\mathcal{D}) & = \int d\boldsymbol{\bar{\varrho}}~\mathcal{P}(\bar{\rho}|\boldsymbol{\bar{\varrho}}) \mathcal{P}(\boldsymbol{\bar{\varrho}}|\mathcal{D}) \\
    & = \frac{1}{P(\boldsymbol{\varrho)}}
    \int~d\boldsymbol{\bar{\varrho}}~\mathcal{P}(\bar{\rho},\boldsymbol{\bar{\varrho}})\, \mathcal{P}(\boldsymbol{\varrho}|\mathcal{D})\, , 
\end{split}
\end{equation}
where $\mathcal{P}(\boldsymbol{\varrho}|\mathcal{D})$ corresponds to observational noise, possibly Gaussian centered in $\boldsymbol{\bar{\varrho}}$ with variance $\sigma_\epsilon^2$. Using the same notation as before, we have $\mathcal{P}(\bar{\rho},\boldsymbol{\bar{\varrho}}) \sim \mathcal{N}(0, \tilde{\Sigma})$ where
\begin{equation}
\label{eq: NNGPCov}
    \tilde{\Sigma} =
    \begin{bmatrix}
    \Sigma(\bar{C},\bar{C}) & \Sigma(\bar{C}, \boldsymbol{\mathcal{C}}) \\
     \Sigma(\boldsymbol{\mathcal{C}}, \bar{C}) & \Sigma(\boldsymbol{\mathcal{C}}, \boldsymbol{\mathcal{C}}) \\
    \end{bmatrix}\, .
\end{equation}
Here, the block structure indicates the division between training and test set. By performing the Gaussian integral over $\boldsymbol{\bar{\varrho}}$, we find $\mathcal{P}(\bar{\rho}|\mathcal{D}) \sim \mathcal{N}(\tilde\mu, \tilde{\Sigma})$ where
\begin{equation}
\begin{split}
\label{eq: GPmusigma}
    & \tilde\mu(\bar{C}) = g(\bar{C},\boldsymbol{\mathcal{C}})~ \boldsymbol{\varrho}\, ,\\&
    \Tilde{\Sigma}(\bar{C},\bar{C}) = \Sigma(\bar{C},\bar{C}) - g(\bar{C},\boldsymbol{\mathcal{C}}) \Sigma(\boldsymbol{\mathcal{C}}, \bar{C})\, ,
\end{split}
\end{equation}
where we introduced the NNGP coefficients  
\begin{equation}
\label{eq: CoeffNNGP}
    g(\bar{C},\boldsymbol{\mathcal{C}})=\Sigma(\bar{C}, \boldsymbol{\mathcal{C}}) \Sigma^{-1}(\boldsymbol{\mathcal{C}}, \boldsymbol{\mathcal{C}}) + \sigma_\epsilon^2\mathds{1},
\end{equation}
which depend on the training data $\boldsymbol{\mathcal{C}}$ and the test input $\bar{C}$. The tilde in Eq.~\eqref{eq: GPmusigma} denotes the posterior mean and covariance. This represents the exact result from Bayesian inference associated to an infinitely wide NN. The kernel $\Sigma$, which is needed in order to compute these Bayesian results, depends on the depth of the NN and can be computed recursively,  see e.g. Refs.~\cite{lee2018deep,lee2019wide} and Appendix~\ref{formulas_sigma_theta} for a short summary.

Moreover, in the large width limit, the NTK at initialization becomes deterministic, 
\begin{equation}
    \label{eq:NTKasymptotic}
    \lim_{n\to\infty} \hat{\Theta}_0 = \Theta_0\, ,
\end{equation}
as summarised in Appendix~\ref{formulas_sigma_theta}, see e.g. Ref.~\cite{jacot2018neural} for more detail. In this limit, the NTK is just an explicit constant kernel, depending only on the depth, activation functions and parameter initialization variance.

\subsection{Interplay between linear and infinite width limit}
\label{Interplay}
In this paragraph we review the relation between the linear and infinite width limit in the case of gradient descent training~\cite{lee2019wide}.

A simplified training is obtained by freezing the hidden layers parameters at their initialization values, i.e. $\theta_t^{(\ell)}=\theta_0^{(\ell)}$, for $\ell<L$, while we optimize the output layer parameters, i.e. $\theta_t^{(L)}$. In this case, even if the hidden layers contain activation functions that generate non-linearities, the optimization only involves the parameters of  the output layer, which is simply a linear combination of the outputs of layer $L-1$. This corresponds to the optimization of the linear approximation of the neural network. Thus, the original network and its linearization are identical and Eq.~\eqref{eq: LinearSol} holds.
Training only the output layer, from the recursive formula of $\Theta^{(L)}_0$ shown in Appendix~\ref{formulas_sigma_theta}, it is clear that $\Theta^{(L)}_0 \to \Sigma^{(L)}$, i.e. it is just the covariance of the output. Thus, by substituting in Eq.~\eqref{eq: muLin} we recover the posterior $\tilde\mu(C)$ in Eq.~\eqref{eq: GPmusigma} in the limit $t\to\infty$. In particular, it can be demonstrated that, for any $t$, the output when training only the $L$-th layer converges, in the infinite width limit, to a Gaussian distribution having expectation value and variance given by 
\begin{equation}
\begin{split}
\label{eq: GPLastLayer}
    \mathbb{E}[\bar{\rho}_t(\bar{C})] &= \Sigma(\bar{C}, \boldsymbol{\mathcal{C}}) \Sigma^{-1}(\boldsymbol{\mathcal{C}}, \boldsymbol{\mathcal{C}})
    (\mathcal{I} - e^{-\eta \Sigma(\boldsymbol{\mathcal{C}}, \boldsymbol{\mathcal{C}}) t}) \boldsymbol{\varrho},\\& = g_t(\bar{C}, \boldsymbol{\mathcal{C}}) \boldsymbol{\varrho}\, ,
\end{split}
\end{equation}
and
\begin{equation}
\begin{split}
    \mathrm{Var}[\bar{\rho}_t(\bar{C})] = &\Sigma(\bar{C},\bar{C}) - \Sigma(\bar{C}, \boldsymbol{\mathcal{C}}) \Sigma^{-1}(\boldsymbol{\mathcal{C}}, \boldsymbol{\mathcal{C}})\times  \\& \times (\mathcal{I} - e^{-2 \eta \Sigma(\boldsymbol{\mathcal{C}}, \boldsymbol{\mathcal{C}}) t})\Sigma(\bar{C}, \boldsymbol{\mathcal{C}})^T\, .
\end{split}
\end{equation}
In the asymptotic limit $t\rightarrow \infty$ we find 
\begin{equation}
g_t(\bar{C}, \boldsymbol{\mathcal{C}}) \rightarrow g(\bar{C}, \boldsymbol{\mathcal{C}}).
\end{equation}
Thus, Eq.~\eqref{eq: GPLastLayer} is identical to the posterior of a GP, i.e. Eq.~\eqref{eq: GPmusigma}. Thus, the NNGP corresponds to the situation in which we only train the output layer of a infinite NN for an infinite time. 

Let us now consider the more general situation in which the gradient descent training optimises the parameters in all the layers of the neural network. In the limit of infinite width, the NTK can be analytically computed by using the recursive relations in Appendix~\ref{formulas_sigma_theta}. 
Furthermore, Ref.~\citep{lee2019wide} shows that, in the limit of infinite width, the dynamics of the gradient descent is equivalent to the one of a linear model, provided the learning rate satisfies $\eta<\eta_{\mathrm{critical}}= 2(\lambda_{\mathrm{min}}(\Theta)+\lambda_{\mathrm{max}}(\Theta))^{-1}$, where $\lambda_{\mathrm{min},\mathrm{max}}$ are respectively the minimum and maximum eigenvalue of the NTK $\Theta$. This means that in the large width limit the dynamics of the original neural network falls automatically into its linearized dynamics regime. Thus, the output is just $\bar{\rho}^{\mathrm{lin}}(\bar{C})$ which is again Gaussian distributed with mean and covariance given by
\begin{equation}
\label{eq: GDEq1}
    \tilde\mu(\bar{C}) = \textsl{g}_t(\bar{C},\boldsymbol{\mathcal{C}})\boldsymbol{\varrho}\, ,
\end{equation} 
and
\begin{equation}
\begin{split}
\label{eq: GDEq2}
    \tilde\Sigma(\bar{C},\bar{C}')  = & \Sigma(\bar{C},\bar{C}') + \textsl{g}_t(\bar{C},\boldsymbol{\mathcal{C}}) \Sigma(\boldsymbol{\mathcal{C}},\boldsymbol{\mathcal{C}})  \textsl{g}_t^T(\bar{C}',\boldsymbol{\mathcal{C}})  \\& - \left[\textsl{g}_t(\bar{C},\boldsymbol{\mathcal{C}})  \Sigma(\boldsymbol{\mathcal{C}},\bar{C}') +h.c. 
    \right]
\end{split}\end{equation}
where we introduced the NTK-GP coefficients
\begin{equation}\label{eq: CoeffNTK-GP}
\textsl{g}_t= \Theta_0(\bar{C},\boldsymbol{\mathcal{C}}) \Theta_0^{-1}(\boldsymbol{\mathcal{C}},\boldsymbol{\mathcal{C}})\big(\mathcal{I}-e^{-\eta \Theta_0 t}\big),
\end{equation}
which depend on the training time that can easily be sent to infinity as in the NNGP case. Here, $\Theta_0$ is the deterministic NTK, see Appendix~\ref{formulas_sigma_theta}.

This limit situation is often called the NTK-GP. In this case, differently from the NNGP case, the output Gaussian distribution \textit{cannot} be interpreted as a posterior of a Bayesian probabilistic model, it is the result of an actual minimization process. Even though the NTK-GP reduces to the NNGP in the linear limit, the two procedures do not agree when all the parameters of the NN are trained. 

\subsection{Relation with Backus Gilbert techniques}\label{Equivalence}
We have discussed two different GPs that emerge from the asymptotic behaviours of NNs: NNGP and NTK-GP. 

In Appendix~\ref{BG review} we recall the equivalence between BG techniques and GPs. In the following, we refer to them as BG-GP, and we refer the reader to Refs.~\cite{10.1093/gji/ggz520,ExtendedTwistedMassCollaborationETMC:2022sta,DelDebbio:2023pyg} for more details. In all cases the central value for the reconstructed function is computed as a linear combination of inputs with some coefficients (cf. Eqs.~\eqref{eq: CoeffNNGP},\eqref{eq: CoeffNTK-GP} and~\eqref{eq: BGGPCoeff}). However, BG-GP and NNGP, while they both have a clear Bayesian interpretation, differ in how their mean values are computed (cf. Eqs.~\eqref{eq: GPmusigma} and~\eqref{eq:BGGPmu}). In the BG-GP case, the coefficients directly weigh time slices of the test correlation function. In NNGP, the spectral function corresponding to the input correlator $\bar{C}$ is a weighted combination of the training spectral functions. The dependence on $\bar{C}$ is solely in the coefficients derived from the covariance matrix~\eqref{eq: NNGPCov}, and distinct from~\eqref{eq: BGGPCov} used for BG-GP. Notably, in NNGP, the reconstructed solution contains covariances between NN outputs, while in BG-GP, it captures correlations between different time slices of the inputs and the output. In the NNGP the latter correlations  are captured by the architecture of the net. This is what makes these two approaches distinct each other. 

Finally, let us observe once again that the NTK-GP does not have a direct Bayesian interpretation,  and seems to be unrelated to the other approaches already used in literature, in particular the BG-GP approach.

\section{Numerical experiments}\label{Experiments}
The relation between NNs and GPs can be used to investigate whether using a full and arbitrarily complex NN would imply advantages in the resolution of convolutional inverse problems. If we send the width of the NN to infinity, the non-linearities generated by the parameters of the NN become negligible, and the statistical distribution of the output of the net tends to a GP. The question is how much the introduction of these non-linearities is relevant in solving convolutional inverse problems. 

In this Section, we carry out a simple numerical experiment in order to compare the outcomes obtained using the various NN limits. We focus on cases where the spectral density is analytically known, specifically, the quantum harmonic oscillator simulated on the lattice trough Monte Carlo techniques, see Appendix~\ref{AppendixA} for details. This application offers the advantage of quantifying the discrepancies between the results obtained using different techniques and the true spectral densities, which are known in the case of the harmonic oscillator.

The numerical experiments are performed on the test data of the correlation function of $y^3$, where $y$ is the coordinate of the harmonic oscillator, for which the analytical result is reported in Eq.~\eqref{eq: True_CorrFunc}. The spectral density is extracted numerically using all the different techniques discussed in the previous Section, namely the NNGP (discussed in Paragraph~\ref{LargeWidth}), the NTK-GP (discussed in Paragraph~\ref{Interplay}) and a finite width NN. 
For the NN we consider a Multilayer Linear Perceptron (MLP) with 4 hidden layers having 16 neurons each. The NN is trained using MSE loss functions and GD training. The training set consists of correlators, $\boldsymbol{\mathcal{C}}$, and their correspondent spectral functions, $\boldsymbol{\varrho}$, computed at discrete points. The input has the dimension of the number of independent temporal points of the correlation function, in our case 100. All the three methods are trained using the same training set, $\{\boldsymbol{\mathcal{C}},\boldsymbol{\varrho}\}$, consisting of mock spectral functions built as a superposition of Gaussian distributions with fixed $\sigma \sim 0.01$ from which the correspondent correlation functions are computed. Further details are given in Appendix~\ref{TrainingApp}. The uncertainties are taken into account by a bootstrap sampling on the configurations of the correlation functions, as previously done in Refs.~\cite{Bonanno:2023ljc,Bonanno:2023thi}. Then, the final result corresponds to the mean value and the variance on the samples of the outcome of the different methods.

\begin{figure}
    \begin{center}
    \includegraphics[scale=0.6]{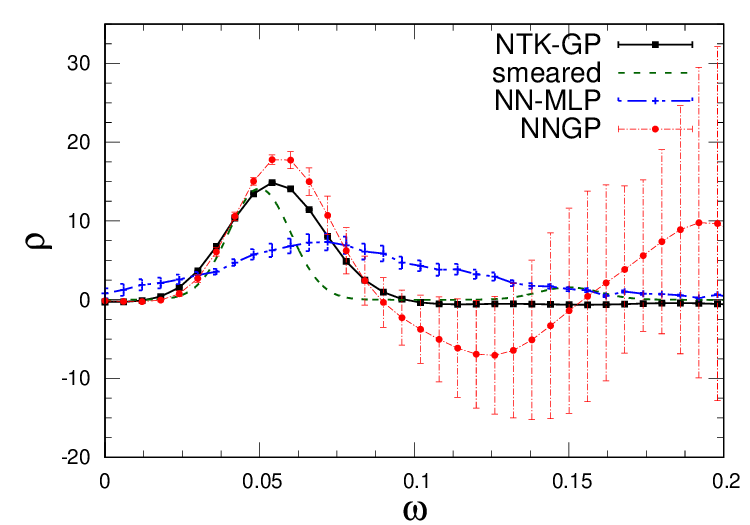}     
    \end{center}
    \caption{\small{Comparison of the results obtained using a neural network and its asymptotic limits corresponding to GPs. The dashed green line represents smeared spectral density, smeared with a Gaussian width $\sigma=0.01$. The three methods are trained using the training set described in Appendix~\ref{TrainingApp} with Gaussian bumps at fixed value of $\sigma=0.01$. The final uncertainties are obtained by bootstrap resampling. In the NTK-GP case they are smaller than the symbols.}}
    \label{Comparison_MT}
\end{figure}

In Fig.~\ref{Comparison_MT}, we show the results obtained using the three different techniques. The plot shows that, for the same training set, both GPs outperform the NN outcomes. It is interesting to focus on the comparison between the two GPs. The NTK-GP results seem to have better controlled uncertainties in contrast to NNGP, where errors explode as $\omega$ increases. The two approaches yield different results, even statistically, due to their distinct limits and approaches. In the NNGP case, there is no GD training on the network parameters (except for the last layer), but a Bayesian training approach. For the NTK-GP, the neural network is fully trained with GD, in the limit where the number of parameters in each layer tends to infinity, with no valid Bayesian interpretation. One possibility is that GD training, compared to Bayesian training, may suppress fluctuations leading to larger errors in the case of the NNGP, and as also observed in the BG-GP case. The small uncertainties of the NTK-GP, not visible in the plot, could be a consequence of the stability of the method. However, since the resulting uncertainties are extremely small, the NNGP should be considered more accurate. Since the statistical uncertainties are all obtained using bootstrap techniques, the different statistical behaviours call for further analysis. Finally, using our training set, all the algorithms are not able to resolve the second, much smaller, peak.

\begin{figure}
    \centering
    \includegraphics[scale=0.6]{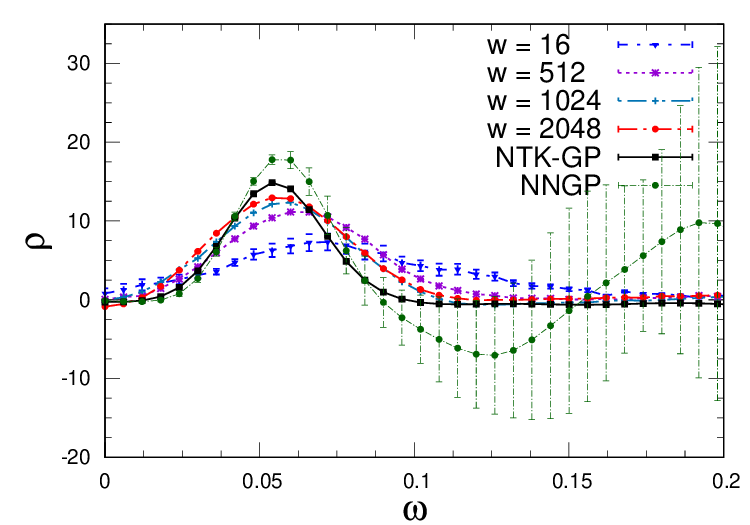}
    \caption{\textit{The figure shows numerically the large width limit of the neural network. Our results clearly show that in the large width limit, $w \rightarrow \infty$, the outcome of the neural network tends to the GPs' ones.}}
    \label{LargeW}
\end{figure}

In Figure~\ref{LargeW}, we show the large width limit of the NN compared with the results obtained using the related GPs. It can be observed that as we increase the number of neurons for each layer the NN results approach the GPs results. This indicates that the finite-width NN accuracy tends to the GPs one as its width is increased.

\section{Conclusions}
We conducted a numerical study on the effectiveness of NN techniques for solving inverse convolution problems. Specifically, drawing on results from computer science theory, we highlighted the asymptotic limits of the NNs. These limits correspond to two different types of GPs,  the NNGP and the NTK-GP, to be added to the already known BG-GP.

The first scenario occurs when the NN has an infinite number of untrained parameters, randomly initialized, resulting in the loss of all non-linearities in those parameters. In this limit the NN defines a Gaussian prior, while the posterior is evaluated analytically from Bayes theorem. In the second scenario, GD training is applied to all layers. At the end of the training, the NNs are normally distributed, but the posterior distribution does not have the covariance expected from Bayes theorem. Nevertheless, even in this case, the NN converges to its linear approximation in the limit of large width. The mean values and covariances of both GPs can be analytically computed as a function of training time.

We considered a case whose inverse problem solution is analytically known, i.e. the quantum harmonic oscillator. Comparing, the results obtained with NNGP to those of a finite-width NN, in the case of a MLP architecture, we showed that NNGP outperforms the latter. In particular, the finite-width NN seems to approach the NNGP accuracy as the layer width is increased. This has already been observed for the MNIST and CIFAR-10 dataset in~\cite{lee2018deep}. Thus, our results, applied in the specific case of convolution inverse problems, seem to confirm a result already established on the two datasets above that are considered important benchmarks in the computer science community. Furthermore, we compared the finite NN outcome with the NTK-GP. The latter, in this case, also proves to perform better compared to the finite-width neural network. 

Our results suggest that the non-linearities introduced by the finite-width NN, which are neglected for NNGP and NTK-GP, may not be necessary in solving this specific inverse problem. Further studies are needed in order to quantify this result and understand in more detail the dynamics of trained NNs. 

At the very least, these two new GPs emerging from the asymptotic behaviour of NNs offer the possibility of investigating inverse problems resolution from a different perspective. This is an approach resulting from the merge of GPs and NN methods, in which the GP basic idea remains but it is equipped with a training process. In particular, the NTK-GP offers a novel perspective compared to established methods in the literature. Furthermore, as underlined in Section~\ref{Equivalence}, they are decoupled with respect the classic GPs approach since they contain the training process, not present in the GPs often used in the literature, which, in principle, could be always improved. 

In any case, our results, which cannot be generalised without extensive, dedicated investigations, beg for a more detailed understanding of the training process for fully connected NNs and other architectures.

Finally, since the computation of the GPs’ output is analytically solvable, the large time output is straightforward achievable by using the limit $t\to\infty$ in the analytical formulas. However, solutions based on finite-width NNs rely on stopping conditions for the training, which never reaches the absolute minimum of the loss function. A quantitative understanding of the stopping criteria is another avenue that requires dedicated studies beyond the scope of this paper.

Let us conclude by underling that the results and conclusion presented in this paper are obtained in the case of a fully connected NN (FCN) and should be extended in the future to other architectures, e.g. convolutional NNs. In Ref.~\cite{novak2018bayesian}, the authors extend the idea of the large-width limit valid for a FCN, to the limit of large number of channels of a CNN. They show that, also in this case, a CNN with many channels corresponds to an NNGP. A careful tuning of the CNN trained with a \textit{stochastic} GD is necessary to obtain results that significantly outperform the correspondent NNGP, at least without pooling. These results call a further investigation in this direction.

\section*{Acknowledgements} It is a pleasure to thank Giuseppe Clemente, Massimo D'Elia, Alessandro De Santis, Alessandro Lupo and Nazario Tantalo for precious discussions and suggestions. M.N. thanks the Higgs Centre for Theoretical Physics of the University of Edinburgh for the hospitality during the first part of the period of this work.   

\newpage
\bibliography{biblio}

\appendix
\section*{SUPPLEMENTARY MATERIAL} 
\section{Backus-Gilbert inversion method}\label{BG review}
In Backus-Gilbert (BG) techniques one looks for approximate solutions of the inverse problem~\eqref{eq: SpectralRelation} as a smeared version of the true spectral function
\begin{equation}
\bar{\rho}(\bar{\omega}) = \int_0^\infty d\omega \Delta(\omega, \bar{\omega})\rho(\omega),
\end{equation}
where $\Delta(\omega, \bar{\omega})$ is a pseudo-gaussian smearing function defined as a linear combination of the basis function in Eq~\eqref{eq: SpectralRelation}
\begin{equation}\label{eq: Smearing}
\Delta(\omega, \bar{\omega}) = \sum_{i=0}^{N_\tau} g_\tau(\bar{\omega})b(\omega,\tau),
\end{equation}
being $N_\tau$ the total number of temporal data.

The idea of Backus-Gilbert techniques, based on Tikhonov regularization methods, is all contained in how the coefficients $g_t$, which define the shape of the function $\Delta$, are determined. The original proposal~\citep{Backus_Gilbert} was to minimise a functional defined as
\begin{equation}\label{eq: Functional}
F[g_\tau] = (1-\lambda)A[g_\tau] + \lambda B[g_\tau],
\end{equation}
where the first term quantifies the width $\sigma$ of the smearing function while the second one the magnitude of the statistical uncertainties related to the final result. The outcome of this method returns a smeared version of the true solution convoluted with a pseudo-gaussian. The width of this Gaussian is determined through the minimization of the functional~\eqref{eq: Functional}, representing a trade-off, regulated by $\lambda \in [0,1]$, between the smearing function's width and the uncertainties magnitude. In absence of uncertainties only the width is minimised. Thus, as more data is used, a tighter width can be reached and the smearing optimised. The ideal case would correspond to the case in which we have infinite data such that $\sigma \rightarrow 0$. 

This method presents a challenge when attempting to estimate a quantity on the lattice. The issue arises because different temporal extensions are typically employed, leading to a variation in the available data with the lattice spacing. As the optimization of the smearing function's width relies on the number of data points used for the minimization, this results in differently smeared outcomes for various time extensions. Consequently, the continuum limit becomes not well-defined. In~\cite{Hansen:2019idp}, a modification of the BG approach is proposed to overcome this issue. In this case the width $\sigma$ of the smearing function is an input of the inversion algorithm. In particular, the $A$ functional does not quantify anymore the width $\sigma$ of the smearing function. Differently, once we fix a target smearing function, the functional quantifies the deviation between the smearing function in Eq.~\eqref{eq: Smearing} and the selected target one. Thus, minimising the functional $A$ would correspond to taking the coefficients $g_\tau$ such that the smearing function~\eqref{eq: Smearing} approximates as better as possible the target one. If the reconstruction is reasonable, this procedure makes it possible to smear consistently the outcome of different lattice spacings and, thus, performing safely the continuum limit. Furthermore, at fixed lattice spacing, it is also possible to study the $\sigma$ dependence of the result and eventually extract predictions for $\sigma \to 0$.

The correspondence between BG techniques and a GPs has been shown in Refs.~\cite{10.1093/gji/ggz520,ExtendedTwistedMassCollaborationETMC:2022sta,DelDebbio:2023pyg}. In particular, the starting point is the extended covariance matrix  
\begin{equation}\label{eq: BGGPCov}
\tilde{\Sigma}_{BG-GP} =
\begin{bmatrix}
  \Sigma\big(C(t), C(t')\big) & \Sigma\big(C(t), \rho(\omega^*)\big) \\
  \Sigma\big(\rho(\omega^*), C(t)\big) & \Sigma\big(\rho(\omega^*),\rho(\omega^*)\big) \\
\end{bmatrix},
\end{equation}
where $\omega^*$ is the point in which we want to extract the correlation function $\rho$, while $C$ is the correspondent correlation function. Note that the term $\Sigma\big(C(t), C(t')\big)$ here represents the covariance between the different points of the same correlation function, while $\Sigma\big(\rho(\omega^*),\rho(\omega^*)\big)$ is just a scalar representing the variance of the output quantity. Thus, using this relation we find that the output $\rho(\omega^*)$ is a Gaussian distribution centred in
\begin{equation}\begin{split}\label{eq:BGGPmu}
\rho(\omega^*)  = \sum_{t,t'} \Sigma\big(C(t'), \rho(\omega^*)\big)^T \frac{1}{\Sigma\big(C(t'), C(t)\big)} C(t) \\=  \sum_t q(t,\omega^*) C(t),
\end{split}\end{equation}
with variance
\begin{equation}\begin{split}\label{eq:BGGPsigma}
\sigma(\omega^*&,\omega^*) =  \Sigma\big(\rho(\omega^*),\rho(\omega^*)\big)+\\& - \Sigma\big(C(t'), \rho(\omega^*)\big)^T \frac{1}{\Sigma\big(C(t'), C(t)\big)} \Sigma\big(C(t), \rho(\omega^*)\big) \\ &
= \Sigma\big(\rho(\omega^*),\rho(\omega^*)\big)  - q(t,\omega^*) \Sigma\big(C(t), \rho(\omega^*)\big),
\end{split}\end{equation}
where we introduced the coefficients 
\begin{equation}\label{eq: BGGPCoeff}
q(t,\omega^*)  = \Sigma\big(C(t'), \rho(\omega^*)\big)^T \Sigma\big(C(t'), C(t)\big)^{-1}.
\end{equation}
The HLT Backus Gilbert method~\cite{Hansen:2019idp} corresponds to the specific situation in which the off-diagonal correlations are taken equal to zero~\cite{ExtendedTwistedMassCollaborationETMC:2022sta,DelDebbio:2023pyg}. Thus, it is contained in this definition as well.

\section{Neural Networks basics and Gradient Descent training}
\label{NN_Intro}
A feedforward Neural Network (NN) consists of multiple layers of artificial neurons. Each 
neuron takes one or more inputs and produces a single output, which is then fed forward to 
the next layer of neurons. The number of neurons in layer $\ell$ is denoted $n_\ell$, and 
the layers are numbered from 0 (the input layer) to $L$ (the output layer). The input layer 
receives the {\em input data}, and the output layer produces the final output of the 
network. 
The neurons in the hidden layers use activation functions to transform the weighted sum of 
their inputs -- called the preactivation function of the neuron -- into a non linear 
output. The weights of the connections between neurons in 
each layer are learned during training using an optimization algorithm like back 
propagation.

Let us consider the $n_0$-dimensional input 
\begin{equation}
\vec{x} = \begin{pmatrix}x_1\\x_2\\... \\x_{n_0}\end{pmatrix}\, .
\end{equation}
The preactivation function for the neurons in the first layer is computed 
by multiplying the input vector by the weight matrix and summing a bias vector. 
The output of the first hidden layer is
\begin{equation}
\rho^{(1)}_i(\vec{x}) = f\left(W_{ij}^{(1)} x_j + b^{(1)}_i\right)\, , \quad i=1,...,n_1\, , 
\end{equation}
where $W^{(1)}$ is an $n_1\times n_0$ weight matrix, while $b^{(1)}$ is an 
$n_1$-dimensional bias vector. The index $j$ is summed over and runs from 1 to $n_0$. 
The function $f$ is the activation function, which 
introduces non-linearities between the different layers. Without a non-linear activation 
function, the network would be limited to linear transformations of the input data, which 
would make it unable to effectively model complex relationships between inputs and outputs. 
Different activation functions have different properties that make them more or less 
suitable for different tasks. We discuss different choices later. The output of the first 
layer is then propagated as the input to the second layer, and the procedure is iterated
for each layer of the NN, 
\begin{equation}
\label{eq:PreActivDef}
\begin{split}
    &\rho^{(\ell+1)}_i(\vec{x}) = f\left(W_{ij}^{(\ell+1)} \rho^{(\ell)}_j(\vec{x}) 
    + b^{(\ell+1)}_i\right)\\
    &\phantom{\rho^{(\ell+1)}_i(\vec{x})} = 
        f\left(\phi^{(\ell+1)}_i(\vec{x})\right)\, ,\\
    &\quad  i=1,...,n_{\ell+1}\, ,
\end{split}
\end{equation}
where in the second line, we introduced the preactivation function $\phi^{(\ell+1)}_i$. 
We refer to the set of {\em all}\ weights and biases as the parameters of the NN, and 
denote them collectively by $\theta$.
Clearly, this recursion relation yields preactivation functions in the deeper layer 
that are nonlinear functions of the parameters of the NN.
Note that in Eq.~\eqref{eq:PreActivDef}, the explicit dependence on the NN parameters 
is omitted for clarity. At times, we will reinstate the explicit dependence and write $\phi^{(\ell)}_i(\vec{x}; \theta)$.
The NN's weights and biases are 
initialized according to some probability density, and subsequent training adjusts 
them to minimize a given loss function. The training phase of a machine learning algorithm 
involves exposing the model, defined by the NN architecture and activation function, 
to a labeled dataset where both inputs and output are known. 
The NN 'learns' to make predictions by adjusting its parameters 
based on the provided data and the minimization of the loss function $\mathcal{L}$. During 
training, the model's parameters are updated iteratively using optimization techniques like 
Gradient Descent (GD) to minimize the loss and improve the predictive accuracy of the 
model. 

Gradient Descent, a widely employed optimization algorithm in neural networks, operates as follows. It computes the gradient of a selected loss function with respect to the NN's parameters. This gradient identifies the direction of steepest ascent in the loss landscape. To minimize the loss, the algorithm updates the parameters by moving in the opposite direction of the gradient. This iterative process continues until convergence or until a predefined stopping criterion is satisfied, resulting in the optimization of the NN's performance. In subsequent phases, such as validation and testing, the trained model is evaluated on new, unseen data to assess its generalization performance. The validation phase helps fine-tuning the hyperparameters and detect overfitting, while the testing phase provides an estimate of how well the model will perform in real-world scenarios.
Once a satisfactory model is obtained, it can be deployed for making predictions on new, 
unseen data, thereby applying the learned patterns to practical tasks or decision-making.

During the training, the NN's output depends on the training time, discarding the
activation function on the output layer, the output of the NN is simply the 
preactivation function of the last layer, $\phi^{(L)}_t$, where we have added a continuous index $t$ that represents the training time. At $t=0$, the NN yields the outcome obtained with the values of the parameters at the initialization, while at subsequent times the value of the parameters is dictated by the training. The weights have the following indices
\begin{equation}
    W = W^{(\ell)}_{t,ij}
\end{equation}
where $\ell \in [0,L]$ identifies the layer, $t$ defines the time of training, and the 
indices $\{i,j\}$ label the matrix elements. The same is valid for the bias vectors, whose 
size is dictated by the number of neurons of the $\ell$-th layer, namely
\begin{equation}
    b = b^{(\ell)}_{t,i}\, , \quad i=1, \ldots, n_\ell\, .
\end{equation}
We denote the set of all the NN parameters (weights and biases) at training 
time $t$ by $\theta_t$, so that $\theta_0$ is the set of parameters at initialization. 
Following the conventions introduced above, the output of the final layer at training 
time $t$ is 
\begin{equation}
    \label{eq:ThetaTimeDep}
    \phi^{(L)}_{t,i}(\vec{x}) = \phi^{(L)}_{i}(\vec{x},\theta_t)\, ,
        \quad i=1,\ldots,n_L\, .   
\end{equation}
Note that the dependence on the training time comes entirely from the parameters 
$\theta_t$. Finally, we call $\mathcal{L}$ the empirical loss function minimized by the 
algorithm to fix the parameters.

Using the Gradient Descent algorithm, the time evolution of $\theta$ and $\phi^{(L)}$ 
is dictated by the following differential equations: 
\begin{equation}\begin{split}\label{eq: GradDesc}
& \dot{\theta}_t= -\eta \sum_{\alpha',i} 
    \nabla_{\theta_t} \phi^{(L)}_{t,i}(\vec{x}_\alpha)\,
    \partial_{\phi^{(L)}_{t,i}(\vec{x}_\alpha)} \mathcal{L}\, ,\\
& \dot{\phi}^{(L)}_{t,i}(x_\alpha) = 
    \nabla_{\theta_t} \phi^{(L)}_{t,i}(\vec{x}_\alpha)\, \dot{\theta}_t \\
& \phantom{\dot{\phi^{(L)}_{t,i}(x_\alpha)}} =     
    -\eta  \sum_{\alpha',j} \hat{\Theta}_{t,ij}(\vec{x}_\alpha, \vec{x}_{\alpha'}) \,
        \partial_{\phi^{(L)}_{t,j}(\vec{x}_{\alpha'})} \mathcal{L}\, .
\end{split}
\end{equation}
The indices $\alpha$ and $\alpha'$ run over the set of training samples, which we denote
$A$. Therefore, the number of training samples is the cardinality of $A$, $|A|$, and 
$\alpha,\alpha'=1, \dots, |A|$.
From the second equation, it is clear that if the parameters do not change in time, 
the output of the NN also remains unchanged.
Here, $\eta$ is the so called learning rate, an hyperparameter that determines how much the 
model's weights are updated in each iteration during the training process. The learning 
rate controls the size of the steps the gradient descent algorithm takes in the opposite 
direction of the loss function's gradient. A learning rate that is too small can make 
training very slow, while a learning rate that is too large can lead to oscillations or 
divergences in the optimization process. Therefore, choosing an appropriate learning rate 
is crucial to ensure effective and stable training of the neural network. 

The quantity $\nabla_{\theta_t} \phi^{(L)}_{t,i}(\vec{x}_\alpha)$, introduced e.g. in 
Ref.~\cite{jacot2018neural}, represents the gradient of the output of the NN with respect 
to its parameters, while $\partial_{\phi^{(L)}_{t,i}(\vec{x}_\alpha)}\mathcal{L}$ is the gradient 
of the loss function with respect to the output of the NN for a given input example 
$x_\alpha$ in the training set. Finally, the matrix $\hat{\Theta}=\hat{\Theta}_{t,ij}(\vec{x}_{\alpha}, \vec{x}_{\alpha'})$ is just the product of two gradients
\begin{equation}
\begin{split}
    \label{eq:NTKdef}
    \hat{\Theta}_{t,ij}(\vec{x}_{\alpha}, \vec{x}_{\alpha'}) & = 
        \nabla_{\theta_t} \phi^{(L)}_{t,i}(\vec{x}_\alpha) \cdot 
        \nabla_{\theta_t} \phi^{(L)}_{t,j}(\vec{x}_{\alpha'})\, ,
\end{split}
\end{equation}
where we recall that $i,j=1,\ldots,n_L$. The matrix $\hat{\Theta}$ is called the
Neural Tangent Kernel (NTK) and plays a major role in describing the training process in what follows. 

\section{Computing the Kernel for NNGP and NKT}
\label{formulas_sigma_theta}

In the large width limit, we can compute analytically the neural tangent kernel $\Theta$ associated with
the NTK-GP and the kernel $\Sigma$ associated with the NNGP. In this procedure, we follow the Refs. 
\cite{lee2018deep, lee2019wide, jacot2018neural} in which $\Theta$ and $\Sigma$ are calculated with recursive functions.

Let us introduce the kernel for the preactivation functions in layer $\ell$,
\begin{align}
    \Sigma^{(\ell)}(\bar{C}_{\delta_1},\bar{C}_{\delta_2})_{ij} &= \lim_{n
    \to \infty} \mathbb{E}\bigg[ \bar{\rho}^{(\ell)}_i(\bar{C}_{\delta_1}) \bar{\rho}^{(\ell)}_j(\bar{C}_{\delta_2}) \bigg] \\
    &= \delta_{ij} 
    \tilde\Sigma^{(\ell)}_{\delta_1\delta_2}\, .
\end{align}
The recursion relation for $\tilde \Sigma(\bar{C}, \bar{C}')$ yields:
\begin{align}
    \tilde \Sigma^{(\ell)}&(\bar{C}, \bar{C}')  
    \nonumber \\
    =& \sigma_w^2 {\cal T} \Biggl(\begin{pmatrix}
        \tilde \Sigma^{(\ell-1)}(\bar{C}, \bar{C}) & \tilde \Sigma^{(\ell-1)}(\bar{C}, \bar{C}') \\
        \tilde \Sigma^{(\ell-1)}(\bar{C}', \bar{C}) & \tilde \Sigma^{(\ell-1)}(\bar{C}',\bar{C}')
    \end{pmatrix} \Biggl) \nonumber \\
    &+ \sigma_b^2 \, , \\
    \tilde \Sigma^{(1)} &(\bar{C}, \bar{C}') = \sigma_w \frac{1}{n^{(0)}} \bar{C}^T \cdot \bar{C}' + \sigma_b^2 \, , 
\end{align}
where $\sigma_w \,, \,\, \sigma_b$ are the weights and biases variances at initialization. 

Similarly, if we consider the NTK at layer $\ell$, 
the recursion relation reads
\begin{align}
\label{eq: RecursiveNTK}
    \tilde \Theta^{(\ell)}& (\bar{C}, \bar{C}') \nonumber \\
    =& \tilde \Sigma^{(\ell)} (\bar{C}, \bar{C}') + \sigma_w^2 \tilde \Theta^{(\ell-1)} (\bar{C}, \bar{C}') \nonumber \\ 
    & \times {\cal T}' \Biggl(\begin{pmatrix}
        \tilde \Sigma^{(\ell-1)}(\bar{C}, \bar{C}) & \tilde \Sigma^{(\ell-1)}(\bar{C}, \bar{C}') \\
        \tilde \Sigma^{(\ell-1)}(\bar{C}', \bar{C}) & \tilde \Sigma^{(\ell-1)}(\bar{C}', \bar{C}')
    \end{pmatrix} \Biggl) \, , \\
    \tilde \Theta^{(1)} &(\bar{C}, \bar{C}')  = \tilde \Sigma^{(1)} (\bar{C}, \bar{C}') \, . 
\end{align}
Finally, the functions $\cal T$ and $\cal T'$ depend by the choice of the activation function, see e.g. Ref.~\cite{Cho09}.

\section{Training set details}\label{TrainingApp}
As a training set for our analysis, we consider input-target pairs consisting of correlators, $C(\tau)$, and their correspondent spectral functions, $\rho(\omega)$. Thus the training data set is given by $\mathcal{D}=\big(\{\boldsymbol{\mathcal{C}}\},\{\boldsymbol{\varrho}\}\big)$ where $\{\boldsymbol{\mathcal{C}}\} = \big(\mathcal{C}(\tau_1), \mathcal{C}(\tau_2),..., \mathcal{C}(\tau_{N_t})\big)$, $\{\boldsymbol{\varrho}\}=\big(\boldsymbol{\varrho}(\omega_1),\boldsymbol{\varrho}(\omega_2), \boldsymbol{\varrho}(\omega_{N_\omega})\big)$.

Since we expect that spectral functions are typically composed of a set of peaks and the smearing is performed with Gaussian distributions, we constructed this training dataset by defining $\boldsymbol{\varrho}(\omega)$ as a superposition of Gaussian distributions. In particular, we defined $\boldsymbol{\varrho}(\omega)$ as

\begin{equation}
\boldsymbol{\varrho}(\omega) = \sum_{i=1}^{N_p} \mathcal{A}_i\mathcal{N}_i(\mu_i, \sigma).
\end{equation}

Here, $N_p$ represents the number of peaks, and we randomly generated the values of the means $\mu_i$  and amplitudes $\mathcal{A}_i$ for each peak. The parameter $\sigma$ is fixed to a predetermined smearing value, $\sigma=0.01$, common to all the peaks. This makes the NN able to solve the inverse problem for input correlation functions corresponding to smearing functions smeared with this fixed $\sigma$ value. Finally, using Eq.~\eqref{eq: SpectralRelation}, we compute the correspondent correlation function to build the training set. 

\begin{figure}
    \centering
    \includegraphics[scale=0.4]{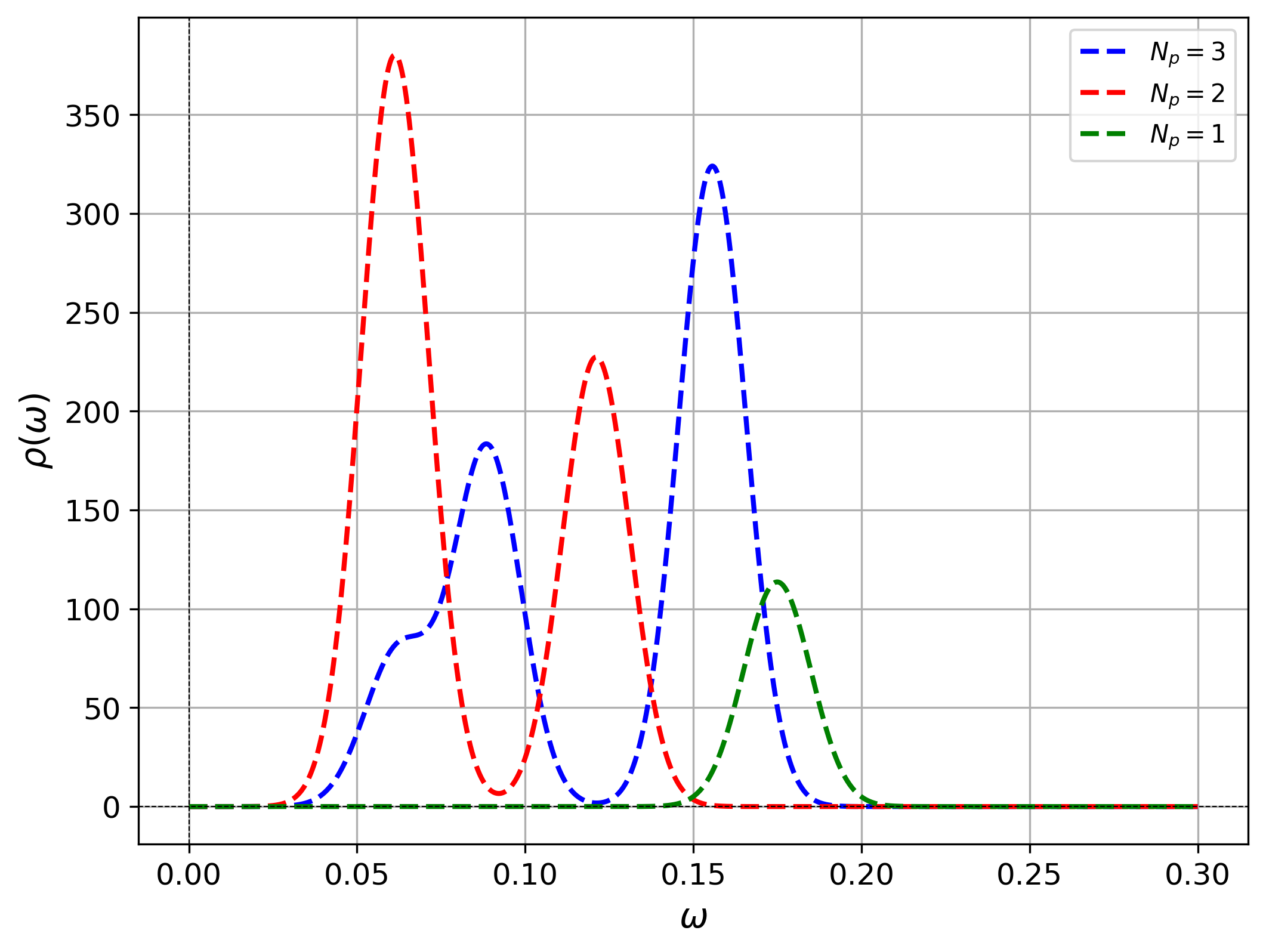}
    \caption{\textit{The figure shows the spectral functions used as training set, generated as a superposition of Gaussian distributions. The value of $\sigma$ is fixed at 0.01, while $\mu$ is randomly generated within the interval [0,0.3].}}
    \label{Training}
\end{figure}

We choose the basis $b(\omega,\tau) = e^{-\omega\tau} + e^{-\omega(L_\tau - \tau) }$, where $L_\tau$ is the time extent of the lattice. In this work $L_\tau=100$, similarly to the test correlation function of the harmonic oscillator discussed in Appendix~\ref{AppendixA}, such that $\tau \in [0,99]$, and $N_p = \{1,2,3\}$, see Fig.~\ref{Training}. As output values $\{\omega\}$ in correspondence of which the spectral functions $\rho(\omega)$ are extracted by the NN, we considered 1000 values of $\omega$ taken in the interval $[0,0.3]$.

In its various limits, the neural network is trained to take as input 100 values of correlators corresponding to different values of $\tau$ and to output 1000 values of the spectral function at the selected values of $\omega$. The Mean Squared Error (MSE) loss function minimizes the difference between the true $\boldsymbol{\varrho}$ and $\bar{\boldsymbol{\rho}}$ predicted by the neural network, denoted as $|\boldsymbol{\varrho} - \bar{\boldsymbol{\rho}}|^2$.

\section{Harmonic Oscillator}\label{AppendixA}
The Euclidean action of the harmonic oscillator reads

\begin{equation}
-\frac{1}{\hbar} S_E = -\frac{1}{\hbar}  \int_0^{\beta\hbar} d\tau \bigg( \frac{m}{2}\bigg(\frac{dx}{d\tau}\bigg)^2 + \frac{m\omega^2}{2}x^2\bigg).
\end{equation}
By discretizing the euclidean time taking $a = \beta\hbar/N$ and taking 
\begin{equation}
\label{eq: y}
y(n) \rightarrow \bigg(\sqrt{\bigg(\frac{\hbar}{m\omega}\bigg)}\bigg)^{-1} x(na), \ \ \ \ \ n = 0, 1, ... , N-1, 
\end{equation}
and using the forward derivative to discretize the free term, then the
Euclidean action assumes the following form
\begin{equation}
\begin{split}
-\frac{1}{\hbar}S_E \rightarrow & -\frac{m\hbar}{2\hbar m \omega} \sum_{n=0}^{N-1} a \bigg( (\frac{y(n+1)-y(n))^2}{a^2} +\omega y(n)^2\bigg) \\& =\sum_{n=0}^{N-1} \bigg(\bigg(\frac{\eta}{2}+\frac{1}{\eta}\bigg)y(n)^2-\frac{1}{\eta}y(n+1)y(n)\bigg)  \\& = -S_D, 
\end{split}\end{equation}
where we defined $\eta = a\omega$. Note that here $\omega$ is not the energy of the system. \\
It's then possible to numerically build a sample which represents a thermodynamic ensemble of the harmonic oscillator where the system parameters are $\eta$ an $N$. Note that the low temperature regime is reached when $\eta N \gg 1$.\\
\subsubsection{Analytical correlation functions}
We want to compute the correlation functions to extract information about the theory. We can write that
\begin{equation}
\begin{split}
\langle x(na = \tau) x(0) \rangle &= \langle 0 | e^\frac{H\tau}{\hbar}x(0)e^\frac{-H\tau}{\hbar}x(0) |0\rangle \\& = \sum_k \langle 0 | e^\frac{H\tau}{\hbar}x(0)e^\frac{-H\tau}{\hbar} |k\rangle \langle m | x(0) | 0 \rangle \\& =  \sum_k e^{-(E_k-E_0)\tau/\hbar}|\langle k | x(0) |0\rangle|^2.
\end{split}\end{equation}
By subtracting the contribution of the eigenstate, we obtain
\begin{equation}
\label{eq: SpectralDec}
\begin{split}
\langle x(\tau) & x(0) \rangle -  |\langle 0 | x(0) | 0 \rangle|^2  = \sum_{k\neq 0}  e^{-(E_k-E_0)a n/\hbar}|\langle k | x(0) |0\rangle|^2  \\& = \int dE \sum_{k\neq 0} \delta(E-a(E_k-E_0)/\hbar) e^{-E a n/\hbar}|\langle k | x(0) |0\rangle|^2\\& = \int dE e^{-E  n} \rho (E), 
\end{split}\end{equation}
where we defined
\begin{equation}
\label{eq: Spectral}
\rho(E) = \sum_{k\neq 0} \delta(E-a(E_k-E_0)/\hbar) |\langle k | x(0) |0\rangle|^2.
\end{equation}
This tells us that the spectral density $\rho(E)$ that we expect to extract using the Backus-Gilbert approach (or similar) should have something which represents a $\delta$-function peaked in the energy gap between each energy eigenstate and the ground state.\\
Let us now just compute the factor $ |\langle k | x(0) |0\rangle|^2$. In particular, not that the same computation could have been performed in the same way also using a different power of $x$. For example, we will consider also the computation in the case of $x^3$. By using the creation and annihilation operators, we have that
\begin{equation}
\begin{split}
& x = \sqrt{\frac{\hbar}{2m \omega}}( a +a^\dagger), \\
&x^2 = \frac{\hbar}{2m \omega}( a^2 + a^{\dagger 2} + 2 a^\dagger a + 1), \\
& x^3 = \bigg(\sqrt{\frac{\hbar}{2m \omega}}\bigg)^3 \bigg(a^3 + a^{\dagger 3} + 3 a^\dagger a^2 + 3a^{\dagger 2} a + 3(a+a^\dagger)\bigg),
\end{split}
\end{equation}
where we used the commutation relation between creation and annihilation operators, namely $a a^\dagger - a^\dagger a = \mathcal{1}  $. At this point, we have that 
\begin{equation}
\begin{split}
& x |0\rangle = \sqrt{\frac{\hbar}{2m \omega}} |1\rangle, \\
& x^2 |0\rangle = \frac{\hbar}{2m \omega} ( |0\rangle +  |2\rangle ),\\
& x^3 |0\rangle = \bigg(\sqrt{\frac{\hbar}{2m \omega}}\bigg)^3 ( |1\rangle + \frac{1}{3}|3\rangle).
\end{split}
\end{equation}
Thus, by substituting in \eqref{eq: Spectral}, we find that in the case of $\langle x(\tau) x(0) \rangle$ the spectral function reads
\begin{equation}
\begin{split}
\rho(E)_x &= \sum_{k\neq 0} \delta(E-a(E_k-E_0)/\hbar) |\langle k | x(0) |0\rangle|^2 \\& = \delta( E - a(E_1-E_0)/\hbar) \sqrt{\frac{\hbar}{2m \omega}} \\& =  \delta (E - a \omega) \sqrt{\frac{\hbar}{2m \omega}} \\& = \delta (E - \eta) \sqrt{\frac{\hbar}{2m \omega}},
\end{split}
\end{equation}
where we recalled $\eta=a\omega$. Thus, we expect a unique peak in correspondence of $E=\eta
$. Differently, in the case of $\langle x^3(\tau) x^3(0) \rangle$ the spectral function is
\begin{equation}
\begin{split}
\rho(E)_{x^3} & = \bigg(\sqrt{\frac{\hbar}{2m \omega}}\bigg)^3 \bigg(\delta( E - a(E_1-E_0)/\hbar) + \\& +\frac{1}{9}\delta( E - (E_3-E_0)) \bigg)   \\&  = \bigg(\sqrt{\frac{\hbar}{2m \omega}}\bigg)^3 \bigg(\delta( E - \eta) +  \frac{1}{9}\delta( E - 3\eta )\bigg).
\end{split}
\end{equation}
Thus, in this case, we expect not only the same peak as before but also a further smaller peak in correspondence of $E=3\eta$.\\
The correlation functions are computed for $y$, defined in Eq. \eqref{eq: y}, instead of $x$. Thus, we have to take into account further factors. The spectral functions are
\begin{equation}\label{eq: True_CorrFunc}
\begin{split}
& \rho(E)_y = \delta (E - \eta) \sqrt{\frac{\hbar}{2m \omega}} \sqrt{\frac{m\omega}{\hbar}} = \frac{1}{\sqrt{2}} \delta (E - \eta),\\ 
& \rho(E)_{y^3} = \bigg(\sqrt{\frac{\hbar}{2m \omega}}\bigg)^3 \bigg(\delta( E - \eta) +  \frac{1}{9}\delta( E - 3\eta )\bigg) \times \\& \times \bigg(\sqrt{\frac{m\omega}{\hbar}} \bigg)^3 = \frac{1}{2^{3/2}}\bigg(\delta( E - \eta) +  \frac{1}{9}\delta( E - 3\eta )\bigg).
\end{split}
\end{equation}

The Harmonic Oscillator has been simulated using the Metropolis algorithm with $\eta=0.05$. Finally, the correlation functions related to the spectral function in Eq.s~\eqref{eq: True_CorrFunc} have been computed with time extent $L_\tau=200$ and periodicity $L_\tau/2$. Thus, the number of independent input points used as input for the NN algorithms is 100.

\end{document}